\documentclass[copyright]{eptcs}

\providecommand{\event}{QFM 2009}
\providecommand{\volume}{??}
\providecommand{\anno}{2009}
\providecommand{\firstpage}{1}
\providecommand{\eid}{??}

\usepackage{amssymb}
\usepackage{epsf}

\def\ms#1{\null\ifmmode\mathord{\mathcode`-="702D\it #1\mathcode`\-="2200}%
	\else$\mathord{\mathcode`-="702D\it #1\mathcode`\-="2200}$\fi}

\newcommand{\cws}[2]
	{\\ \centerline{$#2$} \\[-#1pt]}

\newlength{\spacelen}
\newcommand{\tabspace}[1]
	{\settowidth{\spacelen}{$#1$}
         \hspace*{\spacelen}}

\newcommand{\lap}
	{\mbox{$<$}}
\newcommand{\rap}
	{\mbox{$>$}}

\newcommand{\lsp}
	{[ \! [}
\newcommand{\rsp}
	{] \! ]}

\newcommand{\lmp}
	{\{ \! | \,}
\newcommand{\rmp}
	{\, | \! \}}

\newcommand{\cala}
        {\mathcal{A}}

\newcommand{\calc}
        {\mathcal{C}}

\newcommand{\cale}
        {\mathcal{E}}

\newcommand{\calml}
        {\mathcal{ML}}

\newcommand{\calpl}
        {\mathcal{PL}}

\newcommand{\calsc}
        {\mathcal{SC}}

\newcommand{\natns}
	{\mathbb{N}}

\newcommand{\realns}
	{\mathbb{R}}

\newcommand{\procs}
	{\mathbb{P}}

\newcommand{\tests}
	{\mathbb{T}}

\newcommand{\arrow}[2]
        {\, {\auxarrow\limits^{#1}}_{#2} \,}
\newcommand{\auxarrow}
	{\mathop{- \!\! - \!\!\!\! \longrightarrow}}

\newcommand{\vlarrow}[2]
        {\, {\auxvlarrow\limits^{#1}}_{#2} \,}
\newcommand{\auxvlarrow}
        {\mathop{- \hspace{-0.2cm} - \hspace{-0.2cm} - \hspace{-0.2cm}
	- \hspace{-0.2cm} - \hspace{-0.2cm} - \hspace{-0.2cm}
	- \hspace{-0.2cm} - \hspace{-0.2cm} - \hspace{-0.2cm}
	- \hspace{-0.2cm} - \hspace{-0.2cm} - \hspace{-0.2cm}
	- \hspace{-0.2cm} - \hspace{-0.3cm} \longrightarrow}}

\newcommand{\vvlarrow}[2]
        {\, {\auxvvlarrow\limits^{#1}}_{#2} \,}
\newcommand{\auxvvlarrow}
        {\mathop{- \hspace{-0.2cm} - \hspace{-0.2cm} - \hspace{-0.2cm}
        - \hspace{-0.2cm} - \hspace{-0.2cm} - \hspace{-0.2cm}
        - \hspace{-0.2cm} - \hspace{-0.2cm} - \hspace{-0.2cm}
        - \hspace{-0.2cm} - \hspace{-0.2cm} - \hspace{-0.2cm}
        - \hspace{-0.2cm} - \hspace{-0.2cm} - \hspace{-0.2cm}
        - \hspace{-0.2cm} - \hspace{-0.3cm} \longrightarrow}}

\newcommand{\nil}
	{\underline 0}

\newcommand{\sbis}[1]
	{\sim_{#1}}

\newcommand{\pco}[1]
	{\mathop{\Vert_{#1}}}

\newcommand{\infr}[2]
	{\renewcommand{\arraystretch}{1.5}
	\begin{array}{c}
	#1\\
	\hline
	#2
	\end{array}}

\newcommand{\fullbox}
	{{\mbox{}\nolinebreak\hfill{$\rule{2mm}{2mm}$}}}

\newtheorem{new_theorem}
	{Theorem}[section]

\newtheorem{new_definition}
	[new_theorem]{Definition}

\newtheorem{new_remark}
	[new_theorem]{Remark}

\newtheorem{new_example}
	[new_theorem]{Example}

\newtheorem{new_lemma}
	[new_theorem]{Lemma}

\newtheorem{new_proposition}
	[new_theorem]{Proposition}

\newtheorem{new_corollary}
	[new_theorem]{Corollary}

\newenvironment{definition}
	{\begin{new_definition}\rm}
	{\end{new_definition}}

\newenvironment{proposition}
	{\begin{new_proposition}\rm}
	{\end{new_proposition}}

\newenvironment{theorem}
	{\begin{new_theorem}\rm}
	{\end{new_theorem}}

\title{Markovian Testing Equivalence and \\
       Exponentially Timed Internal Actions}

\author{Marco Bernardo
\institute{Universit\`a di Urbino ``Carlo Bo'' -- Italy \\
	   Istituto di Scienze e Tecnologie dell'Informazione}}

\def\titlerunning{Markovian Testing Equivalence and Exponentially Timed Internal Actions}
\def\authorrunning{M.\ Bernardo}

\begin{document}

\maketitle

\begin{abstract}

\noindent
In the theory of testing for Markovian processes developed so far, exponentially timed internal actions are
not admitted within processes. When present, these actions cannot be abstracted away, because their
execution takes a nonzero amount of time and hence can be observed. On the other hand, they must be
carefully taken into account, in order not to equate processes that are distinguishable from a timing
viewpoint. In this paper, we recast the definition of Markovian testing equivalence in the framework of a
Markovian process calculus including exponentially timed internal actions. Then, we show that the resulting
behavioral equivalence is a congruence, has a sound and complete axiomatization, has a modal logic
characterization, and can be decided in polynomial time.

\end{abstract}

%
%
\section{Introduction}\label{intro}
%
%

Markovian behavioral equivalences are a means to relate and manipulate formal models with an underlying
continuous-time Markov chain (CTMC) semantics. Various proposals have appeared in the literature, which are
extensions of the traditional approaches to the definition of behavioral equivalences. Markovian
bisimilarity~\cite{Hil,Her,BBr} considers two processes to be equivalent whenever they are able to mimic
each other's functional and performance behavior stepwise. Markovian testing equivalence~\cite{Ber2}
considers two processes to be equivalent whenever an external observer is not able to distinguish between
them from a functional or performance viewpoint by interacting with them by means of tests and comparing
their reactions. Markovian trace equivalence~\cite{WBM} considers two processes to be equivalent whenever
they are able to perform computations with the same functional and performance characteristics.

The three Markovian behavioral equivalences mentioned above have different discriminating powers as a
consequence of their different definitions. However, they are all meaningful not only from a functional
standpoint~\cite{Mil,DH,BHR}, but also from a performance standpoint. In fact, Markovian bisimilarity is
known to be in agreement with an exact CTMC-level aggregation called ordinary lumpability~\cite{Hil,Buc},
while Markovian testing and trace equivalences are known to be consistent with a coarser exact CTMC-level
aggregation called T-lumpability~\cite{Ber2,Ber3}.

In this paper, we focus on the treatment of internal actions -- denoted by $\tau$ as usual -- that are
exponentially timed. Unlike internal actions of nondeterministic processes, exponentially timed internal
actions cannot be abstracted away, because their execution takes a nonzero amount of time and hence can be
observed. To be precise, in~\cite{Hil,Bra,BKHW} the issue of abstracting from them has been addressed, but
it remains unclear whether and to what extent abstraction is possible, especially if we want to end up with
a weak Markovian behavioral equivalence that induces a nontrivial, exact CTMC-level aggregation.

The definition of Markovian bisimilarity smoothly includes exponentially timed internal actions, by applying
to them the same exit rate equality check that is applied to exponentially timed visible actions.
Unfortunately, this is not the case with Markovian testing and trace equivalences as witnessed by the theory
developed for them, which does not admit exponentially timed internal actions within processes.

When present, these actions must be carefully taken into account in order not to equate processes that are
distinguishable from a timing viewpoint. As an example, given $\lambda, \mu \in \realns_{> 0}$, processes
``$\lap \tau, \lambda \rap . \nil$'' -- which can only execute an exponentially timed internal action whose
average duration is~$1 / \lambda$ -- and ``$\lap \tau, \mu \rap . \nil$'' -- which can only execute an
exponentially timed internal action whose average duration is $1 / \mu$ -- should not be considered
equivalent if $\lambda > \mu$, as the durations of their actions are sampled from different exponential
probability distributions. Moreover, if they were considered equivalent, then congruence with respect to
alternative and parallel composition would not hold.

With the definition of Markovian testing equivalence given in~\cite{Ber2} -- which compares the
probabilities of passing the same test within the same average time upper bound -- there is no way to
distinguish between the two processes above, as they pass with probability~1 the test comprising only the
success state and with probability~0 any other test, independent of the fixed average time upper bound. In
this paper, we show that a simple way to distinguish between the two processes above consists of imposing an
additional constraint on the length of the successful computations to take into account.

For instance, if we take a test comprising only the success state, the two processes above pass the test
with probability~1 for every average time upper bound if we restrict ourselves to successful computations of
length~0. However, if we move to successful computations of length~1 and we use $1 / \lambda$ as average
time upper bound, it turns out that $\lap \tau, \lambda \rap . \nil$ reaches success with probability~1 --
as it has enough time on average to perform its only action -- whereas $\lap \tau, \mu \rap . \nil$ does not
-- as it has not enough time on average to perform its only action by the deadline. A similar idea applies
to Markovian trace equivalence.

After introducing a Markovian process calculus that includes exponentially timed internal actions
(Sect.~\ref{mpc}), we present a new definition of Markovian testing equivalence that embodies the idea
illustrated above (Sect.~\ref{mte}). Then, we show that $(i)$~it coincides with the equivalence defined
in~\cite{Ber2} when exponentially timed internal actions are absent, $(ii)$~its discriminating power does
not change if we introduce exponentially timed internal actions within tests, and $(iii)$~it inherits the
fully abstract characterization studied in~\cite{Ber2} (Sect.~\ref{mtecondchar}). Furthemore, we show that
it is a congruence with respect to typical dynamic and static operators (Sect.~\ref{mtecongr}) and has a
sound and complete axiomatization for nonrecursive processes (Sect.~\ref{mtescax}), thus overcoming the
limitation to dynamic operators of analogous results contained in~\cite{Ber2}. Finally, we show that it has
a modal logic characterization (Sect.~\ref{mtemlchar}), which is based on the same modal language
as~\cite{Ber4}, and that it can be decided in polynomial time (Sect.~\ref{mteveralg}).


%
%
\section{Markovian Process Calculus}\label{mpc}
%
%

In this section, we present a process calculus in which every action has associated with it a rate that
uniquely identifies its exponentially distributed duration. The definition of the syntax and of the
semantics for the resulting Markovian process calculus -- MPC for short -- is followed by the introduction
of some notations related to process terms and their computations that will be used in the rest of the
paper.

%
\subsection{Durational Actions and Behavioral Operators}
%

In MPC, an exponentially timed action is represented as a pair $\lap a, \lambda \rap$. The first element,
$a$, is the name of the action, which is $\tau$ in the case that the action is internal, otherwise it
belongs to a set $\ms{Name}_{\rm v}$ of visible action names. The second element, $\lambda \in \realns_{>
0}$, is the rate of the exponentially distributed random variable $\ms{RV}$ quantifying the duration of the
action, i.e., $\Pr \{ \ms{RV} \le t \} = 1 - {\rm e}^{- \lambda \cdot t}$ for $t \in \realns_{> 0}$. The
average duration of the action is equal to the reciprocal of its rate, i.e., $1 / \lambda$. If several
exponentially timed actions are enabled, the race policy is adopted: the action that is executed is the
fastest one.

The sojourn time associated with a process term $P$ is thus the minimum of the random variables quantifying
the durations of the exponentially timed actions enabled by $P$. Since the minimum of several exponentially
distributed random variables is exponentially distributed and its rate is the sum of the rates of the
original variables, the sojourn time associated with $P$ is exponentially distributed with rate equal to the
sum of the rates of the actions enabled by $P$. Therefore, the average sojourn time associated with $P$ is
the reciprocal of the sum of the rates of the actions it enables. The probability of executing one of those
actions is given by the action rate divided by the sum of the rates of all the considered actions.

Passive actions of the form $\lap a, *_{w} \rap$ are also included in MPC, where $w \in \realns_{> 0}$ is
the weight of the action. The duration of a passive action is undefined. When several passive actions are
enabled, the reactive preselection policy is adopted. This means that, within every set of enabled passive
actions having the same name, each such action is given an execution probability equal to the action weight
divided by the sum of the weights of all the actions in the set. Instead, the choice among passive actions
having different names is nondeterministic. Likewise, the choice between a passive action and an
exponentially timed action is nondeterministic.

MPC comprises a CSP-like parallel composition operator~\cite{BHR} relying on an asymmetric synchronization
discipline~\cite{BBr}, according to which an exponentially timed action can synchronize only with a passive
action having the same name. In other words, the synchronization between two exponentially timed actions is
forbidden. Following the terminology of~\cite{GSS}, the adopted synchronization discipline mixes generative
and reactive probabilistic aspects. Firstly, among all the enabled exponentially timed actions, the proposal
of an action name is generated after a selection based on the rates of those actions. Secondly, the enabled
passive actions that have the same name as the proposed one react by means of a selection based on their
weights. Thirdly, the exponentially timed action winning the generative selection and the passive action
winning the reactive selection synchronize with each other. The rate of the synchronization is given by the
rate of the selected exponentially timed action multiplied by the execution probability of the selected
passive action, thus complying with the bounded capacity assumption~\cite{Hil}.

We denote by $\ms{Act} = \ms{Name} \times \ms{Rate}$ the set of actions of MPC, where $\ms{Name} =
\ms{Name}_{\rm v} \cup \{ \tau \}$ is the set of action names -- ranged over by $a, b$ -- and $\ms{Rate} =
\realns_{> 0} \cup \{ *_{w} \mid w \in \realns_{> 0} \}$ is the set of action rates -- ranged over by
$\tilde\lambda, \tilde\mu$. We then denote by $\ms{Relab}$ a set of relabeling functions $\varphi :
\ms{Name} \rightarrow \ms{Name}$ that preserve action visibility, i.e., such that $\varphi^{-1}(\tau) = \{
\tau \}$. Finally, we denote by $\ms{Var}$ a set of process variables ranged over by $X, Y$.

	\begin{definition}

The set of process terms of the process language $\calpl$ is generated by the following syntax:
\[\begin{array}{|rcll|}
\hline
P & \!\!\! ::= \!\!\! & \nil & \hspace{0.5cm} \textrm{inactive process} \\
& \!\!\! | \!\!\! & \lap a, \lambda \rap . P & \hspace{0.5cm} \textrm{exponentially timed action prefix} \\
& \!\!\! | \!\!\! & \lap a, *_{w} \rap . P & \hspace{0.5cm} \textrm{passive action prefix} \\
& \!\!\! | \!\!\! & P + P & \hspace{0.5cm} \textrm{alternative composition} \\
& \!\!\! | \!\!\! & P \pco{S} P & \hspace{0.5cm} \textrm{parallel composition} \\
& \!\!\! | \!\!\! & P / H & \hspace{0.5cm} \textrm{hiding} \\
& \!\!\! | \!\!\! & P[\varphi] & \hspace{0.5cm} \textrm{relabeling} \\
& \!\!\! | \!\!\! & X & \hspace{0.5cm} \textrm{process variable} \\
& \!\!\! | \!\!\! & \ms{rec} \, X : P & \hspace{0.5cm} \textrm{recursion} \\
\hline
\end{array}\]
where $a \in \ms{Name}$, $\lambda, w \in \realns_{> 0}$, $S, H \subseteq \ms{Name}_{\rm v}$, $\varphi \in
\ms{Relab}$, and $X \in \ms{Var}$. We denote by $\procs$ the set of closed and guarded process terms of
$\calpl$.
\fullbox

	\end{definition}

%
\subsection{Operational Semantics}
%

The semantics for MPC can be defined in the usual operational style, with an important difference with
respect to the nondeterministic case. A process term like $\lap a, \lambda \rap . \nil + \lap a, \lambda
\rap . \nil$ is not the same as $\lap a, \lambda \rap . \nil$, because the average sojourn time associated
with the latter, i.e., $1 / \lambda$, is twice the average sojourn time associated with the former, i.e., $1
/ (\lambda + \lambda)$. In order to assign distinct semantic models to terms like the two considered above,
we have to take into account the multiplicity of each transition, intended as the number of different proofs
for the transition derivation. The semantic model $\lsp P \rsp$ for a process term $P \in \procs$ is thus a
labeled multitransition system, whose multitransition relation is contained in the smallest multiset of
elements of $\procs \times \ms{Act} \times \procs$ satisfying the operational semantic rules of
Table~\ref{sos} ($\{ \_ \hookrightarrow \_ \}$ denotes syntactical replacement; $\lmp, \rmp$ are multiset
parentheses).

	\begin{table}[p]

\[\begin{array}{|c|}
\hline
(\textsc{Pre}_{1}) \quad {\infr{}{\lap a, \lambda \rap . P \arrow{a, \lambda}{} P}} \hspace{1.5cm}
(\textsc{Pre}_{2}) \quad {\infr{}{\lap a, *_{w} \rap . P \arrow{a, *_{w}}{} P}} \\[0.9cm]
(\textsc{Alt}_{1}) \quad {\infr{P_{1} \arrow{a, \tilde\lambda}{} P'}{P_{1} + P_{2} \arrow{a,
\tilde\lambda}{} P'}} \hspace{1.5cm}
(\textsc{Alt}_{2}) \quad {\infr{P_{2} \arrow{a, \tilde\lambda}{} P'}{P_{1} + P_{2} \arrow{a,
\tilde\lambda}{} P'}} \\[0.9cm]
(\textsc{Par}_{1}) \quad {\infr{P_{1} \arrow{a, \tilde\lambda}{} P'_{1} \hspace{0.8cm} a \notin S}{P_{1}
\pco{S} P_{2} \arrow{a, \tilde\lambda}{} P'_{1} \pco{S} P_{2}}} \hspace{1.5cm}
(\textsc{Par}_{2}) \quad {\infr{P_{2} \arrow{a, \tilde\lambda}{} P'_{2} \hspace{0.8cm} a \notin S}{P_{1}
\pco{S} P_{2} \arrow{a, \tilde\lambda}{} P_{1} \pco{S} P'_{2}}} \\[0.9cm]
(\textsc{Syn}_{1}) \quad {\infr{P_{1} \arrow{a, \lambda}{} P'_{1} \hspace{0.8cm} P_{2} \arrow{a, *_{w}}{}
P'_{2} \hspace{0.8cm} a \in S}{P_{1} \pco{S} P_{2} \vlarrow{a, \lambda \cdot {w \over \ms{weight}(P_{2},
a)}}{} P'_{1} \pco{S} P'_{2}}} \\[0.9cm]
(\textsc{Syn}_{2}) \quad {\infr{P_{1} \arrow{a, *_{w}}{} P'_{1} \hspace{0.8cm} P_{2} \arrow{a, \lambda}{}
P'_{2} \hspace{0.8cm} a \in S}{P_{1} \pco{S} P_{2} \vlarrow{a, \lambda \cdot {w \over \ms{weight}(P_{1},
a)}}{} P'_{1} \pco{S} P'_{2}}} \\[0.9cm]
(\textsc{Syn}_{3}) \quad {\infr{P_{1} \arrow{a, *_{w_{1}}}{} P'_{1} \hspace{0.8cm} P_{2} \arrow{a,
*_{w_{2}}}{} P'_{2} \hspace{0.8cm} a \in S}{P_{1} \pco{S} P_{2} \vvlarrow{a, *_{\ms{norm}(w_{1}, w_{2}, a,
P_{1}, P_{2})}}{} P'_{1} \pco{S} P'_{2}}} \\[0.9cm]
(\textsc{Hid}_{1}) \quad {\infr{P \arrow{a, \tilde\lambda}{} P' \hspace{0.8cm} a \in H}{P / H \arrow{\tau,
\tilde\lambda}{} P' / H}} \hspace{1.5cm}
(\textsc{Hid}_{2}) \quad {\infr{P \arrow{a, \tilde\lambda}{} P' \hspace{0.8cm} a \notin H} {P / H \arrow{a,
\tilde\lambda}{} P' / H}} \\[0.9cm]
(\textsc{Rel}) \quad {\infr{P \arrow{a, \tilde\lambda}{} P'}{P[\varphi] \arrow{\varphi(a), \tilde\lambda}{}
P'[\varphi]}} \\[0.9cm]
(\textsc{Rec}) \quad {\infr{P \{ \ms{rec} \, X : P \hookrightarrow X \} \arrow{a, \tilde\lambda}{}
P'}{\ms{rec} \, X : P \arrow{a, \tilde\lambda}{} P'}} \\[0.9cm]
\hline
\\[-0.2cm]
\ms{weight}(P, a) \: = \: \sum \lmp w \in \realns_{> 0} \mid \exists P' \in \procs \ldotp P \arrow{a,
*_{w}}{} P' \rmp \\[0.3cm]
\ms{norm}(w_{1}, w_{2}, a, P_{1}, P_{2}) \: = \: {w_{1} \over \ms{weight}(P_{1}, a)} \cdot {w_{2} \over
\ms{weight}(P_{2}, a)} \cdot (\ms{weight}(P_{1}, a) + \ms{weight}(P_{2}, a)) \\[0.4cm]
\hline
\end{array}\]

\caption{Operational semantic rules for MPC}\label{sos}

	\end{table}

We observe that exponential distributions fit well with the interleaving view of parallel composition. Due
to their memoryless property, the execution of an exponentially timed action can be thought of as being
started in the last state in which the action is enabled. Due to their infinite support, the probability
that two concurrent exponentially timed actions terminate simultaneously is zero.

The CTMC underlying a process term $P \in \procs$ can be derived from $\lsp P \rsp$ iff this labeled
multitransition system has no passive transitions, in which case we say that $P$ is performance closed. We
denote by $\procs_{\rm pc}$ the set of performance closed process terms of $\procs$.

%
\subsection{Exit Rates of Process Terms}
%

The exit rate of a process term $P \in \procs$ is the rate at which $P$ can execute actions of a certain
name $a \in \ms{Name}$ that lead to a certain destination $D \subseteq \procs$ and is given by the sum of
the rates of those actions due to the race policy. We consider a two-level definition of exit rate, with
level~$0$ corresponding to exponentially timed actions and level~$-1$ corresponding to passive actions:
\[\begin{array}{|c|}
\hline
\ms{rate}_{\rm e}(P, a, l, D) \: = \: \left\{ \begin{array}{ll}
\sum \lmp \lambda \in \realns_{> 0} \mid \exists P' \in D \ldotp P \arrow{a, \lambda}{} P' \rmp &
\hspace{0.5cm} \textrm{if $l = 0$} \\
\sum \lmp w \in \realns_{> 0} \mid \exists P' \in D \ldotp P \arrow{a, *_{w}}{} P' \rmp & \hspace{0.5cm}
\textrm{if $l = -1$} \\
\end{array}\right. \\
\hline
\end{array}\]
where each summation is taken to be zero whenever its multiset is empty.

By summing up the rates of all the actions of a certain level $l$ that $P$ can execute, we obtain the total
exit rate of $P$ at level $l$:
\[\begin{array}{|c|}
\hline
\ms{rate}_{\rm t}(P, l) \: = \: \sum\limits_{a \in \ms{Name}} \ms{rate}_{\rm o}(P, a, l) \\
\hline
\end{array}\]
where:
\[\begin{array}{|c|}
\hline
\ms{rate}_{\rm o}(P, a, l) \: = \: \ms{rate}_{\rm e}(P, a, l, \procs) \\
\hline
\end{array}\]
is the overall exit rate of $P$ with respect to $a$ at level $l$.

If $P$ is performance closed, then $\ms{rate}_{\rm t}(P, 0)$ coincides with the reciprocal of the average
sojourn time associated with $P$. Instead, $\ms{rate}_{\rm o}(P, a, -1)$ coincides with $\ms{weight}(P, a)$.

%
\subsection{Probability and Duration of Computations}\label{probdurcomp}
%

A computation of a process term $P \in \procs$ is a sequence of transitions that can be executed starting
from $P$. The length of a computation is given by the number of transitions occurring in it. We denote by
$\calc_{\rm f}(P)$ the multiset of finite-length computations of $P$. We say that two distinct computations
are independent of each other if neither is a proper prefix of the other one. In the following, we
concentrate on finite multisets of independent, finite-length computations. Below we define the probability
and the duration of a computation $c \in \calc_{\rm f}(P)$ for $P \in \procs_{\rm pc}$, using $\_ \circ \_$
for sequence concatenation and $| \_ |$ for sequence length.

The probability of executing $c$ is the product of the execution probabilities of the transitions of $c$:
\[\begin{array}{|c|}
\hline
\ms{prob}(c) \: = \: \left\{ \begin{array}{ll}
1 &
\hspace{0.5cm} \textrm{if $|c| = 0$} \\
{\lambda \over \ms{rate}_{\rm t}(P, 0)} \cdot \ms{prob}(c') &
\hspace{0.5cm} \textrm{if $c \equiv P \arrow{a, \lambda}{} c'$} \\
\end{array} \right. \\
\hline
\end{array}\]
We also define the probability of executing a computation in $C \subseteq \calc_{\rm f}(P)$ as:
\[\begin{array}{|c|}
\hline
\ms{prob}(C) \: = \: \sum\limits_{c \in C} \ms{prob}(c) \\
\hline
\end{array}\]
whenever $C$ is finite and all of its computations are independent of each other.

The stepwise average duration of $c$ is the sequence of average sojourn times in the states traversed by
$c$:
\[\begin{array}{|c|}
\hline
\ms{time}_{\rm a}(c) \: = \: \left\{ \begin{array}{ll}
\varepsilon &
\hspace{0.5cm} \textrm{if $|c| = 0$} \\
{1 \over \ms{rate}_{\rm t}(P, 0)} \circ \ms{time}_{\rm a}(c') &
\hspace{0.5cm} \textrm{if $c \equiv P \arrow{a, \lambda}{} c'$} \\
\end{array} \right. \\
\hline
\end{array}\]
where $\varepsilon$ is the empty stepwise average duration. We also define the multiset of computations in
$C \subseteq \calc_{\rm f}(P)$ whose stepwise average duration is not greater than $\theta \in (\realns_{>
0})^{*}$ as:
\[\begin{array}{|l|}
\hline
C_{\le \theta} \: = \: \lmp c \in C \mid |c| \le |\theta| \land \forall i = 1, \dots, |c| \ldotp
\ms{time}_{\rm a}(c)[i] \le \theta[i] \rmp \\
\hline
\end{array}\]
Moreover, we denote by $C^{l}$ the multiset of computations in $C \subseteq \calc_{\rm f}(P)$ whose length
is equal to $l \in \natns$.

We conclude by observing that the average duration of a finite-length computation has been defined as the
sequence of average sojourn times in the states traversed by the computation. The same quantity could have
been defined as the sum of the same basic ingredients, but this would not have been appropriate as explained
in~\cite{WBM,Ber2}.

%
%
\section{Redefining Markovian Testing Equivalence}\label{mte}
%
%

The basic idea behind testing equivalence is to infer information about the behavior of process terms by
interacting with them by means of tests and comparing their reactions. In a Markovian setting, we are not
only interested in verifying whether tests are passed or not, but also in measuring the probability with
which they are passed and the time taken to pass them. Therefore, we have to restrict ourselves to
$\procs_{\rm pc}$.

As in the nondeterministic setting, the most convenient way to represent a test is through a process term,
which interacts with any process term under test by means of a parallel composition operator that enforces
synchronization on the set $\ms{Name}_{\rm v}$ of all visible action names. Due to the adoption of an
asymmetric synchronization discipline, a test can comprise only passive visible actions, so that the
composite term inherits performance closure from the process term under test.

From a testing viewpoint, in any of its states a process term under test generates the proposal of an action
to be executed by means of a race among the exponentially timed actions enabled in that state. If the name
of the proposed action is $\tau$, then the process term advances by itself. Otherwise, the test either
reacts by participating in the interaction with the process term through a passive action having the same
name as the proposed exponentially timed action, or blocks the interaction if it has no passive actions with
the proposed name.

Markovian testing equivalence relies on comparing the process term probabilities of performing successful
test-driven computations within arbitrary sequences of average amounts of time. Due to the presence of these
average time upper bounds, for the test representation we can restrict ourselves to nonrecursive process
terms. In other words, the expressiveness provided by finite-state labeled multitransition systems with an
acyclic structure is enough for tests.

In order not to interfere with the quantitative aspects of the behavior of process terms under test, we
avoid the introduction of a success action $\omega$. The successful completion of a test is formalized in
the text syntax by replacing $\nil$ with a zeroary operator s denoting a success state. Ambiguous tests
including several summands among which at least one equal to s are avoided through a two-level syntax.

	\begin{definition}

The set $\tests_{\rm R}$ of reactive tests is generated by the following syntax:
\[\begin{array}{|rcl|}
\hline
T & \!\!\! ::= \!\!\! & \textrm{s} \mid T' \\
T' & \!\!\! ::= \!\!\! & \lap a, *_{w} \rap . T \mid T' + T' \\
\hline
\end{array}\]
where $a \in \ms{Name}_{\rm v}$ and $w \in \realns_{> 0}$.
\fullbox

	\end{definition}

	\begin{definition}

Let $P \in \procs_{\rm pc}$ and $T \in \tests_{\rm R}$. The interaction system of $P$ and $T$ is process
term $P \pco{\ms{Name}_{\rm v}} T \in \procs_{\rm pc}$ and we say that:

		\begin{itemize}

\item A configuration is a state of $\lsp P \pco{\ms{Name}_{\rm v}} T \rsp$, which is formed by a process
and a test projection.

\item A configuration is successful iff its test projection is s.

\item A test-driven computation is a computation of $\lsp P \pco{\ms{Name}_{\rm v}} T \rsp$.

\item A test-driven computation is successful iff it traverses a successful configuration.

		\end{itemize}

\noindent
We denote by $\calsc(P, T)$ the multiset of successful computations of $P \pco{\ms{Name}_{\rm v}} T$.
\fullbox

	\end{definition}

If a process term $P \in \procs_{\rm pc}$ under test has no exponentially timed $\tau$-actions as it was
in~\cite{Ber2}, then for all reactive tests $T \in \tests_{\rm R}$ it turns out that: $(i)$~all the
computations in $\calsc(P, T)$ have a finite length due to the restrictions imposed on the test syntax;
$(ii)$~all the computations in $\calsc(P, T)$ are independent of each other because of their maximality;
$(iii)$~the multiset $\calsc(P, T)$ is finite because $P$ and $T$ are finitely branching. Thus, all
definitions of Sect.~\ref{probdurcomp} are applicable to $\calsc(P, T)$ and also to $\calsc_{\le \theta}(P,
T)$ for any sequence $\theta \in (\realns_{> 0})^{*}$ of average amounts of time.

In order to cope with the possible presence of exponentially timed $\tau$-actions within $P$ in such a way
that all the properties above hold -- especially independence -- we have to consider subsets of $\calsc_{\le
\theta}(P, T)$ including all successful test-driven computations of the same length. This is also necessary
to distinguish among process terms comprising only exponentially timed $\tau$-actions -- like $\lap \tau,
\lambda \rap . \nil$ and $\lap \tau, \mu \rap . \nil$, with $\lambda > \mu$, mentioned in Sect.~\ref{intro}
-- as there is a single test, s, that those process terms can pass. The only option is to compare them after
executing the same number of $\tau$-actions.

Since no element of $\calsc_{\le \theta}(P, T)$ can be longer than $|\theta|$, we should consider every
possible subset $\calsc_{\le \theta}^{l}(P, T)$ for $0 \le l \le |\theta|$. However, it is enough to
consider $\calsc_{\le \theta}^{|\theta|}(P, T)$, as shorter successful test-driven computations can be taken
into account when imposing prefixes of $\theta$ as average time upper bounds. Therefore, the novelty with
respect to~\cite{Ber2} is simply the presence of the additional constraint $|\theta|$.

	\begin{definition}\label{newmtedef}

Let $P_{1}, P_{2} \in \procs_{\rm pc}$. We say that $P_{1}$ is Markovian testing equivalent to $P_{2}$,
written $P_{1} \sbis{\rm MT} P_{2}$, iff for all reactive tests $T \in \tests_{\rm R}$ and sequences $\theta
\in (\realns_{> 0})^{*}$ of average amounts of time:
\cws{10}{\ms{prob}(\calsc_{\le \theta}^{|\theta|}(P_{1}, T)) \: = \: \ms{prob}(\calsc_{\le
\theta}^{|\theta|}(P_{2}, T))}
\fullbox

	\end{definition}

Note that we have not defined a may equivalence and a must equivalence as in the nondeterministic
case~\cite{DH}. The reason is that in this Markovian framework the possibility and the necessity of passing
a test are not sufficient to discriminate among process terms, as they are qualitative concepts. What we
have considered here is a single quantitative notion given by the probability of passing a test (within an
average time upper bound); hence, the definition of a single equivalence. This quantitative notion subsumes
both the possibility of passing a test -- which can be encoded as the probability of passing the test being
greater than zero -- and the necessity of passing a test -- which can be encoded as the probability of
passing the test being equal to one.

Although we could have defined Markovian testing equivalence as the kernel of a Markovian testing preorder,
this has not been done. The reason is that such a preorder would have boiled down to an equivalence
relation, because for each reactive test passed by $P_{1}$ within $\theta$ with a probability less than the
probability with which $P_{2}$ passes the same test within $\theta$, in general it is possible to find a
dual reactive test for which the relation between the two probabilities is inverted.

Another important difference with respect to the nondeterministic case is that the presence of average time
upper bounds makes it possible to decide whether a test is passed or not even if the process term under test
can execute infinitely many exponentially timed $\tau$-actions. In other words, $\tau$-divergence
does not need to be taken into account.

%
%
\section{Basic Properties and Characterizations}\label{mtecondchar}
%
%

First of all, we observe that, whenever exponentially timed $\tau$-actions are absent, the new Markovian
testing equivalence $\sbis{\rm MT}$ coincides with the old one defined in~\cite{Ber2}, which we denote by
$\sbis{\rm MT, old}$. In the following, we use $\procs_{\rm pc, v}$ to refer to the process terms of
$\procs_{\rm pc}$ that contain no exponentially timed $\tau$-actions.

	\begin{proposition}\label{mteoldnewprop}

Let $P_{1}, P_{2} \in \procs_{\rm pc, v}$. Then $P_{1} \sbis{\rm MT} P_{2} \: \Longleftrightarrow \: P_{1}
\sbis{\rm MT, old} P_{2}$.
\fullbox

	\end{proposition}

Then, we have two alternative characterizations of $\sbis{\rm MT}$, which provide further justifications for
the way in which the equivalence has been defined. The first one establishes that the discriminating power
does not change if we consider a set $\tests_{\rm R, lib}$ of tests with the following more liberal syntax:
\cws{0}{T \: ::= \: \textrm{s} \mid \lap a, *_{w} \rap . T \mid T + T}
provided that by successful configuration we mean a configuration whose test projection includes s as
top-level summand. Let us denote by $\sbis{\rm MT, lib}$ the resulting variant of Markovian testing
equivalence.

	\begin{proposition}\label{mtelibcharprop}

Let $P_{1}, P_{2} \in \procs_{\rm pc}$. Then $P_{1} \sbis{\rm MT, lib} P_{2} \: \Longleftrightarrow \: P_{1}
\sbis{\rm MT} P_{2}$.
\fullbox

	\end{proposition}

The second characterization establishes that the discriminating power does not change if we consider a set
$\tests_{{\rm R}, \tau}$ of tests capable of moving autonomously by executing exponentially timed
$\tau$-actions:
\cws{0}{\begin{array}{rcl}
T & \!\!\! ::= \!\!\! & \textrm{s} \mid T' \\
T' & \!\!\! ::= \!\!\! & \lap a, *_{w} \rap . T \mid \lap \tau, \lambda \rap . T \mid T' + T' \\
\end{array}}
Let us denote by $\sbis{\rm MT, \tau}$ the resulting variant of Markovian testing equivalence.

	\begin{proposition}\label{mtetaucharprop}

Let $P_{1}, P_{2} \in \procs_{\rm pc}$. Then $P_{1} \sbis{\rm MT, \tau} P_{2} \: \Longleftrightarrow \:
P_{1} \sbis{\rm MT} P_{2}$.
\fullbox

	\end{proposition}

Finally, we have two further alternative characterizations of $\sbis{\rm MT}$ coming from~\cite{Ber2}. The
first one establishes that the discriminating power does not change if we consider the (more accurate)
probability distribution of passing tests within arbitrary sequences of amounts of time, rather than the
(easier to work with) probability of passing tests within arbitrary sequences of average amounts of time.

The second characterization fully abstracts from comparing process term behavior in response to tests. This
is achieved by considering traces that are extended at each step with the set of visible action names
permitted by the environment at that step (not to be confused with a ready set). A consequence of the
structure of extended traces is the identification of a set $\tests_{\rm R, c}$ of canonical reactive tests,
which is generated by the following syntax:
\[\begin{array}{|l|}
\hline
T \: ::= \: \textrm{s} \mid \lap a, *_{1} \rap . T + \sum\limits_{b \in \cale - \{ a \}} \hspace{-0.2cm}
\lap b, *_{1} \rap . \lap \textrm{z}, *_{1} \rap . \textrm{s} \\
\hline
\end{array}\]
where $a \in \cale$, $\cale \subseteq \ms{Name}_{\rm v}$ finite, the summation is absent whenever $\cale =
\{ a \}$, and z is a visible action name representing failure that can occur within tests but not within
process terms under test. Similar to the case of probabilistic testing equivalence~\cite{Chr,CDSY}, each of
these canonical reactive tests admits a single computation leading to success, whose intermediate states can
have additional computations each leading to failure in one step. We point out that the canonical reactive
tests are name deterministic, in the sense that the names of the passive actions occurring in any of their
branches are all distinct.

%
%
\section{Congruence Property}\label{mtecongr}
%
%

Markovian testing equivalence is a congruence with respect to all MPC operators. In particular,
unlike~\cite{Ber2}, we have a full congruence result with respect to parallel composition.

	\begin{theorem}\label{mtecongrthm}

Let $P_{1}, P_{2} \in \procs_{\rm pc}$. Whenever $P_{1} \sbis{\rm MT} P_{2}$, then:

		\begin{enumerate}

\item $\lap a, \lambda \rap . P_{1} \sbis{\rm MT} \lap a, \lambda \rap . P_{2}$ for all $\lap a, \lambda
\rap \in \ms{Act}$.

\item $P_{1} + P \sbis{\rm MT} P_{2} + P$ and $P + P_{1} \sbis{\rm MT} P + P_{2}$ for all $P \in \procs_{\rm
pc}$.

\item $P_{1} \pco{S} P \sbis{\rm MT} P_{2} \pco{S} P$ and $P \pco{S} P_{1} \sbis{\rm MT} P \pco{S} P_{2}$
for all $P \in \procs$ and $S \subseteq \ms{Name}_{\rm v}$ s.t.\ $P_{1} \pco{S} P, P_{2} \pco{S} P \in
\procs_{\rm pc}$.

\item $P_{1} / H \sbis{\rm MT} P_{2} / H$ for all $H \subseteq \ms{Name}_{\rm v}$.

\item $P_{1}[\varphi] \sbis{\rm MT} P_{2}[\varphi]$ for all $\varphi \in \ms{Relab}$.
\fullbox

		\end{enumerate}

	\end{theorem}

It is worth stressing that the additional constraint on the length of successful test-driven computations
present in Def.~\ref{newmtedef} is fundamental for achieving congruence with respect to alternative and
parallel composition. As an example, if it were $\lap \tau, \lambda \rap . \nil \sbis{\rm MT} \lap \tau, \mu
\rap . \nil$ for $\lambda > \mu$, then we would have $\lap \tau, \lambda \rap . \nil + \lap a, \gamma \rap .
\nil \not\sbis{\rm MT} \lap \tau, \mu \rap . \nil + \lap a, \gamma \rap . \nil$. In fact, when the average
time upper bound is high enough, the probability of passing $\lap a, *_{1} \rap . \textrm{s}$ is ${\gamma
\over \lambda + \gamma}$ for the first term, whereas it is ${\gamma \over \mu + \gamma}$ for the second
term. We also mention that Props.~\ref{mtelibcharprop} and~\ref{mtetaucharprop} are exploited in the
congruence proof for static operators.

%
%
\section{Sound and Complete Axiomatization}\label{mtescax}
%
%

Markovian testing equivalence has a sound and complete axiomatization over the set $\procs_{\rm pc, nrec}$
of nonrecursive process terms of $\procs_{\rm pc}$, given by the set $\cala_{\rm MT}$ of equational laws of
Table~\ref{mteaxioms}.

	\begin{table}[p]

\[\begin{array}{|crcl|}
\hline
(\cala_{{\rm MT}, 1}) & P_{1} + P_{2} & \!\!\! = \!\!\! & P_{2} + P_{1} \\
(\cala_{{\rm MT}, 2}) & (P_{1} + P_{2}) + P_{3} & \!\!\! = \!\!\! & P_{1} + (P_{2} + P_{3}) \\
(\cala_{{\rm MT}, 3}) & P + \nil & \!\!\! = \!\!\! & P \\[0.5cm]
(\cala_{{\rm MT}, 4}) & \sum\limits_{i \in I} \lap a, \lambda_{i} \rap . \sum\limits_{j \in J_{i}} \lap
b_{i, j}, \mu_{i, j} \rap . P_{i, j} & \!\!\! = \!\!\! & \lap a, \mathop{\rm \Sigma}\limits_{k \in I}
\lambda_{k} \rap . \sum\limits_{i \in I} \sum\limits_{j \in J_{i}} \lap b_{i, j}, {\lambda_{i} \over
\mathop{\rm \Sigma}_{k \in I} \lambda_{k}} \cdot \mu_{i, j} \rap . P_{i, j} \\[0.4cm]
& & & \hspace*{-4.9cm} \textrm{if: $I$ is a finite index set with $|I| \ge 2$;} \\
& & & \hspace*{-4.4cm} \textrm{for all $i \in I$, index set $J_{i}$ is finite and its summation is $\nil$ if
$J_{i} = \emptyset$;} \\
& & & \hspace*{-4.4cm} \textrm{for all $i_{1}, i_{2} \in I$ and $b \in \ms{Name}$:} \\
& \sum\limits_{j \in J_{i_{1}}} \lmp \mu_{i_{1}, j} \mid b_{i_{1}, j} = b \rmp & \!\!\! = \!\!\! &
\sum\limits_{j \in J_{i_{2}}} \lmp \mu_{i_{2}, j} \mid b_{i_{2}, j} = b \rmp \\[0.8cm]
(\cala_{{\rm MT}, 5}) & \hspace{0.1cm} \sum\limits_{i \in I} \lap a_{i}, \tilde\lambda_{i} \rap . P_{i} \,
\pco{S} \, \sum\limits_{j \in J} \lap b_{j}, \tilde\mu_{j} \rap . Q_{j} & \!\!\! = \!\!\! & \\[0.3cm]
& & & \hspace*{-2.3cm} \sum\limits_{k \in I, a_{k} \notin S} \hspace{-0.2cm} \lap a_{k}, \tilde\lambda_{k}
\rap . \left( P_{k} \, \pco{S} \, \sum\limits_{j \in J} \lap b_{j}, \tilde\mu_{j} \rap . Q_{j} \right) \; +
\\[0.3cm]
& & & \hspace*{-2.3cm} \sum\limits_{h \in J, b_{h} \notin S} \hspace{-0.2cm} \lap b_{h}, \tilde\mu_{h} \rap
. \left( \sum\limits_{i \in I} \lap a_{i}, \tilde\lambda_{i} \rap . P_{i} \, \pco{S} \, Q_{h} \right) \; +
\\[0.4cm]
& & & \hspace*{-2.3cm} \sum\limits_{k \in I, a_{k} \in S, \tilde\lambda_{k} \in \realns_{> 0}} \;
\sum\limits_{h \in J, b_{h} = a_{k}, \tilde\mu_{h} = *_{w_{h}}} \hspace{-0.4cm} \lap a_{k},
\tilde\lambda_{k} \cdot {w_{h} \over \ms{weight}(Q, b_{h})} \rap . (P_{k} \, \pco{S} \, Q_{h}) \; +
\\[0.4cm]
& & & \hspace*{-2.3cm} \sum\limits_{h \in J, b_{h} \in S, \tilde\mu_{h} \in \realns_{> 0}} \; \sum\limits_{k
\in I, a_{k} = b_{h}, \tilde\lambda_{k} = *_{v_{k}}} \hspace{-0.4cm} \lap b_{h}, \tilde\mu_{h} \cdot {v_{k}
\over \ms{weight}(P, a_{k})} \rap . (P_{k} \, \pco{S} \, Q_{h}) \; + \\[0.4cm]
& & & \hspace*{-2.3cm} \sum\limits_{k \in I, a_{k} \in S, \tilde\lambda_{k} = *_{v_{k}}} \; \sum\limits_{h
\in J, b_{h} = a_{k}, \tilde\mu_{h} = *_{w_{h}}} \hspace{-0.4cm} \lap a_{k}, *_{\ms{norm}(v_{k}, w_{h},
a_{k}, P, Q)} \rap . (P_{k} \, \pco{S} \, Q_{h}) \\[0.6cm]
(\cala_{{\rm MT}, 6}) & \sum\limits_{i \in I} \lap a_{i}, \tilde\lambda_{i} \rap . P_{i} \, \pco{S} \, \nil
& \!\!\! = \!\!\! & \sum\limits_{k \in I, a_{k} \notin S} \hspace{-0.2cm} \lap a_{k}, \tilde\lambda_{k} \rap
. P_{k} \\[0.4cm]
(\cala_{{\rm MT}, 7}) & \nil \, \pco{S} \, \sum\limits_{j \in J} \lap b_{j}, \tilde\mu_{j} \rap . Q_{j} &
\!\!\! = \!\!\! & \sum\limits_{h \in J, b_{h} \notin S} \hspace{-0.2cm} \lap b_{h}, \tilde\mu_{h} \rap .
Q_{h} \\[0.4cm]
(\cala_{{\rm MT}, 8}) & \nil \, \pco{S} \, \nil & \!\!\! = \!\!\! & \nil \\[0.8cm]
(\cala_{{\rm MT}, 9}) & \nil / H & \!\!\! = \!\!\! & \nil \\
(\cala_{{\rm MT}, 10}) & (\lap a, \tilde\lambda \rap . P) / H & \!\!\! = \!\!\! & \lap \tau, \tilde\lambda
\rap . (P / H) \hspace{2.3cm} \textrm{if $a \in H$} \\
(\cala_{{\rm MT}, 11}) & (\lap a, \tilde\lambda \rap . P) / H & \!\!\! = \!\!\! & \lap a, \tilde\lambda \rap
. (P / H) \hspace{2.3cm} \textrm{if $a \notin H$} \\
(\cala_{{\rm MT}, 12}) & (P_{1} + P_{2}) / H & \!\!\! = \!\!\! & P_{1} / H + P_{2} / H \\[0.5cm]
(\cala_{{\rm MT}, 13}) & \nil[\varphi] & \!\!\! = \!\!\! & \nil \\
(\cala_{{\rm MT}, 14}) & (\lap a, \tilde\lambda \rap . P)[\varphi] & \!\!\! = \!\!\! & \lap \varphi(a),
\tilde\lambda \rap . (P[\varphi]) \\
(\cala_{{\rm MT}, 15}) & (P_{1} + P_{2})[\varphi] & \!\!\! = \!\!\! & P_{1}[\varphi] + P_{2}[\varphi] \\
\hline
\end{array}\]

\caption{Equational laws for $\sbis{\rm MT}$}\label{mteaxioms}

	\end{table}

Apart from the usual laws for the alternative composition operator and for the unary static operators,
unlike the axiomatization of~\cite{Ber2} we now have laws dealing with concurrency. In particular, axiom
$\cala_{{\rm MT}, 5}$ concerning the parallel composition of $P \equiv \sum_{i \in I} \lap a_{i},
\tilde\lambda_{i} \rap . P_{i}$ and $Q \equiv \sum_{j \in J} \lap b_{j}, \tilde\mu_{j} \rap . Q_{j}$ --
where $I$ and $J$ are nonempty finite index sets and each summation on the right-hand side of the axiom is
taken to be $\nil$ whenever its set of summands is empty -- is the expansion law when enforcing
generative-reactive and reactive-reactive synchronizations. This axiom applies to non-performance-closed
process terms too; e.g., the last addendum on its right-hand side is related to reactive-reactive
synchronizations.

Like in~\cite{Ber2}, the law characterizing $\sbis{\rm MT}$ is the axiom schema $\cala_{{\rm MT}, 4}$, which
in turn subsumes the law $\lap a, \lambda_{1} \rap . P + \lap a, \lambda_{2} \rap . P = \lap a, \lambda_{1}
+ \lambda_{2} \rap . P$ characterizing Markovian bisimilarity. The simplest instance of axiom schema
$\cala_{{\rm MT}, 4}$ is depicted below: \\[0.1cm]
\centerline{\epsfbox{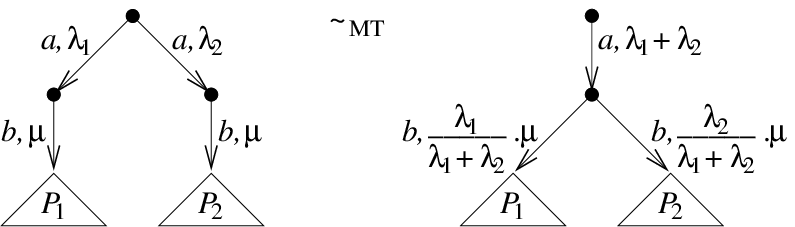}}
As emphasized by the figure above, $\sbis{\rm MT}$ allows choices to be deferred in the case of branches
that start with the same action name (see the two $a$-branches on the left-hand side) and are followed by
sets of actions having the same names and total rates (see $\{ \lap b, \mu \rap \}$ after each of the two
$a$-branches).

	\begin{theorem}\label{mtescaxthm}

Let $P_{1}, P_{2} \in \procs_{\rm pc, nrec}$. Then $\cala_{\rm MT} \vdash P_{1} = P_{2} \:
\Longleftrightarrow \: P_{1} \sbis{\rm MT} P_{2}$.
\fullbox

	\end{theorem}

%
%
\section{Modal Logic Characterization}\label{mtemlchar}
%
%

Markovian testing equivalence has a modal logic characterization that, as in~\cite{Ber4}, is based on a
modal language comprising true, disjunction, and diamond. A constraint is imposed on formulas of the form
$\phi_{1} \lor \phi_{2}$, which does not reduce the expressive power as it is consistent with the
name-deterministic nature of branches within canonical reactive tests (see Sect.~\ref{mtecondchar}).

	\begin{definition}\label{modallanguage}

The set of formulas of the modal language $\calml_{\rm MT}$ is generated by the following syntax:
\[\begin{array}{|rcl|}
\hline
\phi & \!\!\! ::= \!\!\! & \textrm{true} \mid \phi' \\
\phi' & \!\!\! ::= \!\!\! & \langle a \rangle \phi \mid \phi' \lor \phi' \\
\hline
\end{array}\]
where $a \in \ms{Name}_{\rm v}$ and each formula of the form $\phi_{1} \lor \phi_{2}$ satisfies:
\cws{0}{\ms{init}(\phi_{1}) \cap \ms{init}(\phi_{2}) \: = \: \emptyset}
with $\ms{init}(\phi)$ being defined by induction on the syntactical structure of $\phi$ as follows:
\cws{10}{\begin{array}{rcl}
\ms{init}(\textrm{true}) & \!\!\! = \!\!\! & \emptyset \\
\ms{init}(\langle a \rangle \phi) & \!\!\! = \!\!\! & \{ a \} \\
\ms{init}(\phi_{1} \lor \phi_{2}) & \!\!\! = \!\!\! & \ms{init}(\phi_{1}) \cup \ms{init}(\phi_{2}) \\
\end{array}}
\fullbox

	\end{definition}

Probabilistic and temporal information do not decorate any operator of the modal language, but come into
play through a quantitative interpretation function inspired by~\cite{KN} that replaces the usual boolean
satisfaction relation. This interpretation function measures the probability that a process term satisfies a
formula quickly enough on average. The constraint imposed by Def.~\ref{modallanguage} on disjunctions
guarantees that their subformulas exercise independent computations of the process term, thus ensuring the
correct calculation of the probability of satisfying the overall formula. In order to manage exponentially
timed $\tau$-actions, unlike~\cite{Ber4} the length of the computations satisfying the formula has to be
taken into account as well.

	\begin{definition}

The interpretation function $\lsp . \rsp_{\rm MT}^{.}$ of $\calml_{\rm MT}$ over $\procs_{\rm pc} \times
(\realns_{> 0})^{*}$ is defined by letting:
\[\begin{array}{|l|}
\hline
\lsp \phi \rsp_{\rm MT}^{|\theta|} (P, \theta) \: = \: \left\{ \begin{array}{ll}
0 &
\hspace{0.5cm} \textrm{if $|\theta| = 0 \land \phi \not\equiv \textrm{true}$ or} \\
& \tabspace{\hspace{0.5cm} \textrm{if }}
\textrm{$|\theta| > 0 \land \ms{rate}_{\rm o}(P, \ms{init}(\phi) \cup \{ \tau \}, 0) = 0$} \\[0.2cm]
1 &
\hspace{0.5cm} \textrm{if $|\theta| = 0 \land \phi \equiv \textrm{true}$} \\
\end{array} \right. \\
\hline
\end{array}\]
otherwise by induction on the syntactical structure of $\phi$ and on the length of $\theta$ as follows:
\[\begin{array}{|rcl|}
\hline
\lsp \textrm{true} \rsp_{\rm MT}^{|t \circ \theta|} (P, t \circ \theta) & \!\!\! = \!\!\! & \left\{
\begin{array}{ll}
\sum\limits_{P \arrow{\tau, \lambda}{} P'} \hspace{-0.3cm} {\lambda \over \ms{rate}_{\rm o}(P, \tau, 0)}
\cdot \lsp \textrm{true} \rsp_{\rm MT}^{|\theta|} (P', \theta) &
\hspace{0.5cm} \textrm{if ${1 \over \ms{rate}_{\rm o}(P, \tau, 0)} \le t$} \\[0.6cm]
0 &
\hspace{0.5cm} \textrm{if ${1 \over \ms{rate}_{\rm o}(P, \tau, 0)} > t$} \\
\end{array} \right. \\[1.0cm]
\lsp \langle a \rangle \phi \rsp_{\rm MT}^{|t \circ \theta|} (P, t \circ \theta) & \!\!\! = \!\!\! & \left\{
\begin{array}{ll}
\sum\limits_{P \arrow{a, \lambda}{} P'} \hspace{-0.3cm} {\lambda \over \ms{rate}_{\rm o}(P, \{ a, \tau \},
0)} \cdot \lsp \phi \rsp_{\rm MT}^{|\theta|} (P', \theta) \; + & \\
\hspace{0.4cm} \sum\limits_{P \arrow{\tau, \lambda}{} P'} \hspace{-0.3cm} {\lambda \over \ms{rate}_{\rm
o}(P, \{ a, \tau \}, 0)} \cdot \lsp \langle a \rangle \phi \rsp_{\rm MT}^{|\theta|} (P', \theta) &
\hspace{0.5cm} \textrm{if ${1 \over \ms{rate}_{\rm o}(P, \{ a, \tau \}, 0)} \le t$} \\[0.6cm]
0 &
\hspace{0.5cm} \textrm{if ${1 \over \ms{rate}_{\rm o}(P, \{ a, \tau \}, 0)} > t$} \\
\end{array} \right. \\[1.6cm]
\lsp \phi_{1} \lor \phi_{2} \rsp_{\rm MT}^{|t \circ \theta|} (P, t \circ \theta) & \!\!\! = \!\!\! & p_{1}
\cdot \lsp \phi_{1} \rsp_{\rm MT}^{|t_{1} \circ \theta|} (P_{\ms{no-init-\tau}}, t_{1} \circ \theta) + p_{2}
\cdot \lsp \phi_{2} \rsp_{\rm MT}^{|t_{2} \circ \theta|} (P_{\ms{no-init-\tau}}, t_{2} \circ \theta)
\\[0.2cm]
& & \; + \sum\limits_{P \arrow{\tau, \lambda}{} P'} \hspace{-0.3cm} {\lambda \over \ms{rate}_{\rm o}(P,
\ms{init}(\phi_{1} \lor \phi_{2}) \cup \{ \tau \}, 0)} \cdot \lsp \phi_{1} \lor \phi_{2} \rsp_{\rm
MT}^{|\theta|} (P', \theta) \\
\hline
\end{array}\]
where $P_{\ms{no-init-\tau}}$ is $P$ devoid of all of its computations starting with a $\tau$-transition --
which is assumed to be $\nil$ whenever all the computations of $P$ start with a $\tau$-transition -- and for
$j \in \{ 1, 2 \}$:
\cws{10}{\begin{array}{rclcrcl}
p_{j} & \!\!\! = \!\!\! & {\ms{rate}_{\rm o}(P, \ms{init}(\phi_{j}), 0) \over \ms{rate}_{\rm o}(P,
\ms{init}(\phi_{1} \lor \phi_{2}) \cup \{ \tau \}, 0)} & & t_{j} & \!\!\! = \!\!\! & t + ({1 \over
\ms{rate}_{\rm o}(P, \ms{init}(\phi_{j}), 0)} - {1 \over \ms{rate}_{\rm o}(P, \ms{init}(\phi_{1} \lor
\phi_{2}) \cup \{ \tau \}, 0)}) \\
\end{array}}
\fullbox

	\end{definition}

In the definition above, $p_{j}$ represents the probability with which $P$ performs actions whose name is in
$\ms{init}(\phi_{j})$ rather than actions whose name is in $\ms{init}(\phi_{k}) \cup \{ \tau \}$, $k = 3 -
j$, given that $P$ can perform actions whose name is in $\ms{init}(\phi_{1} \lor \phi_{2}) \cup \{ \tau \}$.
These probabilities are used as weights for the correct account of the probabilities with which $P$
satisfies only $\phi_{1}$ or $\phi_{2}$ in the context of the satisfaction of $\phi_{1} \lor \phi_{2}$. If
such weights were omitted, then the fact that $\phi_{1} \lor \phi_{2}$ offers a set of initial actions at
least as large as the ones offered by $\phi_{1}$ alone and by $\phi_{2}$ alone would be ignored, thus
leading to a potential overestimate of the probability of satisfying $\phi_{1} \lor \phi_{2}$.

Similarly, $t_{j}$ represents the extra average time granted to $P$ for satisfying only~$\phi_{j}$. This
extra average time is equal to the difference between the average sojourn time in $P$ when only actions
whose name is in $\ms{init}(\phi_{j})$ are enabled and the average sojourn time in $P$ when also actions
whose name is in $\ms{init}(\phi_{k}) \cup \{ \tau \}$, $k = 3 - j$, are enabled. Since the latter cannot be
greater than the former due to the race policy -- more enabled actions means less time spent on average in a
state -- considering $t$ instead of $t_{j}$ in the satisfaction of $\phi_{j}$ in isolation would lead to a
potential underestimate of the probability of satisfying $\phi_{1} \lor \phi_{2}$ within the given average
time upper bound, as $P$ may satisfy $\phi_{1} \lor \phi_{2}$ within $t \circ \theta$ even if $P$ satisfies
neither $\phi_{1}$ nor $\phi_{2}$ taken in isolation within $t \circ \theta$.

	\begin{theorem}\label{mtemlcharthm}

$P_{1} \sbis{\rm MT} P_{2} \: \Longleftrightarrow \: \forall \phi \in \calml_{\rm MT} \ldotp \forall \theta
\in (\realns_{> 0})^{*} \ldotp \lsp \phi \rsp_{\rm MT}^{|\theta|} (P_{1}, \theta) = \lsp \phi \rsp_{\rm
MT}^{|\theta|} (P_{2}, \theta)$.
\fullbox

	\end{theorem}

%
%
\section{Verification Algorithm}\label{mteveralg}
%
%

Markovian testing equivalence can be decided in polynomial time. The reason is that Markovian testing
equivalence coincides with Markovian ready equivalence and, given two process terms, their underlying CTMCs
in which action names have not been discarded from transition labels are Markovian ready equivalent iff the
corresponding embedded DTMCs in which transitions have been labeled with suitably augmented names are
related by probabilistic ready equivalence. The latter equivalence is decidable in polynomial time~\cite{HT}
through a reworking of the algorithm for probabilistic language equivalence~\cite{Tze}.

Following~\cite{WBM}, the transformation of a name-labeled CTMC into the corresponding embedded name-labeled
DTMC is carried out by simply turning the rate of each transition into the corresponding execution
probability. Then, we need to encode the total exit rate of each state of the original name-labeled CTMC
inside the names of all transitions departing from that state in the associated embedded DTMC.

\bigskip
\noindent
\textbf{Acknowledgment}: This work has been funded by MIUR-PRIN project \textit{PaCo -- Performability-Aware
Computing: Logics, Models, and Languages}.

\bibliographystyle{eptcs}

\end{document}